\documentclass[%
 aip,
 amsmath,amssymb,
reprint,%
]{revtex4-1}

\usepackage{graphicx}
\usepackage{dcolumn}
\usepackage{bm}

\usepackage[utf8]{inputenc}
\usepackage[T1]{fontenc}
\usepackage{mathptmx}
\usepackage{etoolbox}
\usepackage{natbib}
\usepackage{graphicx}
\usepackage{latexsym}
\usepackage{amssymb}
\usepackage{amsmath}
\usepackage{amsfonts}
\usepackage{gensymb}
\usepackage{bm}
\usepackage{hyperref}
\usepackage{braket}
\usepackage{physics}
\usepackage{bbold}
\usepackage{slashed}
\usepackage[dvipsnames]{xcolor}
\usepackage[normalem]{ulem}
\usepackage{siunitx}
\usepackage[version=4]{mhchem}

\makeatletter
\def\@email#1#2{%
 \endgroup
 \patchcmd{\titleblock@produce}
  {\frontmatter@RRAPformat}
  {\frontmatter@RRAPformat{\produce@RRAP{*#1\href{mailto:#2}{#2}}}\frontmatter@RRAPformat}
  {}{}
}%
\makeatother
\begin{document}

\preprint{AIP/123-QED}

\title[MilliKelvin microwave impedance microscopy in a dry dilution refrigerator]{MilliKelvin microwave impedance microscopy in a dry dilution refrigerator }
\author{Leonard Weihao Cao}
\affiliation{Department of Physics, University of California, San Diego, CA 92093, USA \looseness=-1}

\author{Chen Wu}
\affiliation{Department of Physics, University of California, San Diego, CA 92093, USA \looseness=-1}
\author{Rajarshi Bhattacharyya}
\affiliation{Department of Physics, University of California, San Diego, CA 92093, USA \looseness=-1}
\author{Ruolun Zhang}
\affiliation{Department of Physics, University of California, San Diego, CA 92093, USA \looseness=-1}

\author{Monica T. Allen}%
 \email{mtallen@physics.ucsd.edu.}
\affiliation{Department of Physics, University of California, San Diego, CA 92093, USA \looseness=-1}

\date{\today}

\begin{abstract}
Microwave impedance microscopy (MIM) is a near-field imaging technique that has been used to visualize the local conductivity of materials with nanoscale resolution across the GHz regime. 
In recent years, MIM has shown great promise for the investigation of topological states of matter, correlated electronic states and emergent phenomena in quantum materials. To explore these low-energy phenomena, many of which are only detectable in the milliKelvin regime, we have developed a novel low-temperature MIM incorporated into a dilution refrigerator. This setup, which consists of a tuning-fork-based atomic force microscope with microwave reflectometry capabilities, is capable of reaching temperatures down to \SI{70}{mK} during imaging and magnetic fields up to \SI{9}{T}. To test the performance of this microscope, we demonstrate microwave imaging of the conductivity contrast between graphite and silicon dioxide at cryogenic temperatures and discuss the resolution and noise observed in these results. We extend this methodology to visualize edge conduction in Dirac semimetal cadmium arsenide in the quantum Hall regime.

\end{abstract}

\maketitle

\section{\label{sec:level1}Introduction}


\begin{figure*}
\includegraphics[width=0.95\textwidth]{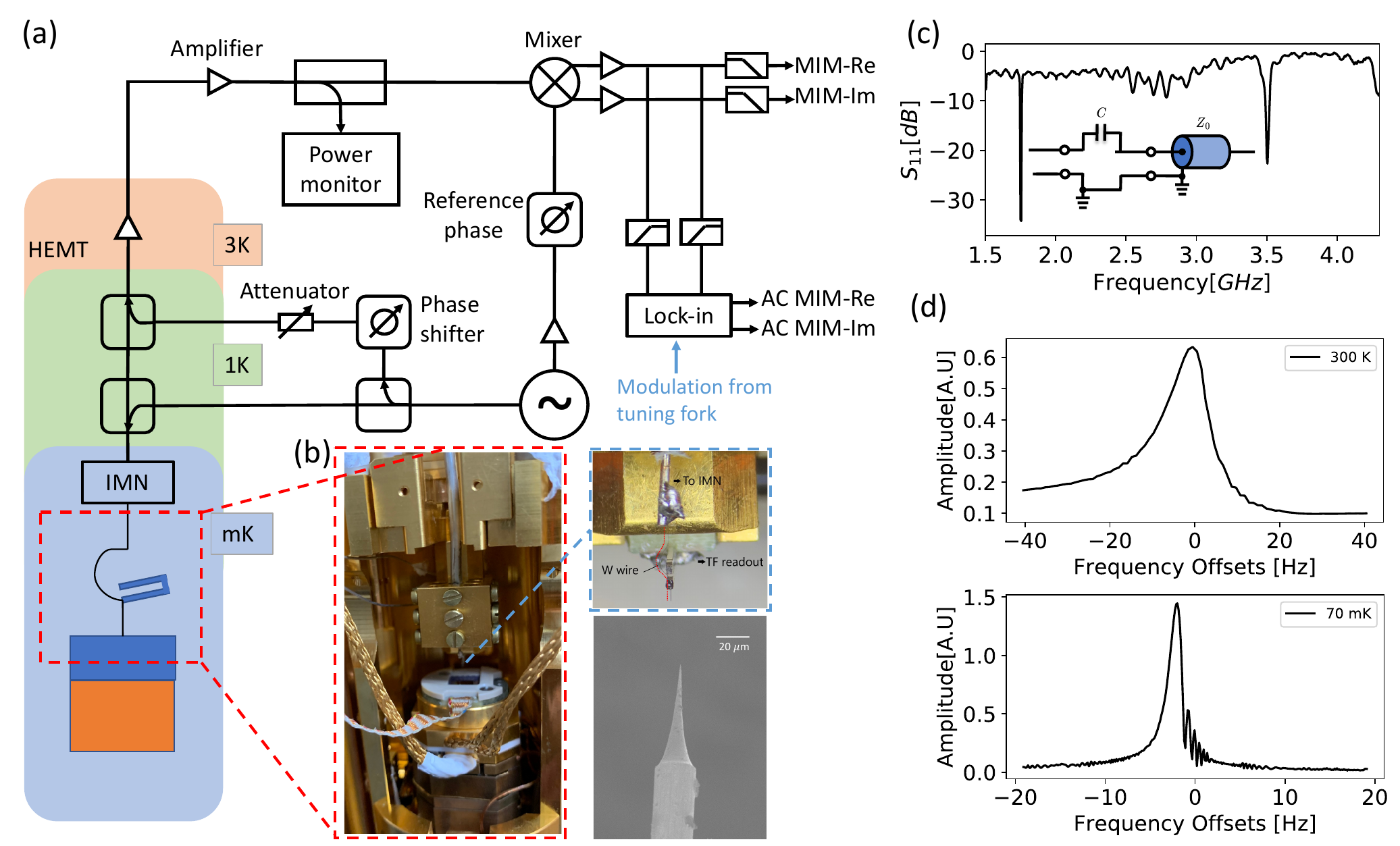}
\caption{\label{fig:schematics} \textbf{Scanning probe microscopy system with combined AFM and MIM readout, integrated into a dilution refrigerator.} 
\textbf{(a)} 
Schematic of the scanning MIM readout electronics and hardware integrated into a dilution refrigerator. 
The shaded regions refer to the different thermal stages inside the fridge. 
\textbf{(b)} \textit{Left panel:} 
Photo of the microscope head, scanners and sample stage (corresponding to the red box in the schematic). 
\textit{Right panels}: Zoomed-in side view of the gold-coated tungsten tip glued onto one prong of the tuning fork, and soldered to the end of the IMN on the other end. The back of the TF is electrically connected for oscillation readout (blue box); Scanning electron microscope image of the etched tungsten tip used for combined AFM and MIM imaging. 
\textbf{(c)} Plot of the reflect microwave power $S_{11}$ of the impedance matching network (IMN) shows the fundamental resonance at the \SI{1.8}{GHz}. \textit{Inset:} Circuit diagram of the IMN with a \SI{0.2}{pF} capacitor with \SI{5}{cm} coax connected in series.
\textbf{(d)} Plots of the oscillation amplitude of the tuning fork as a function of frequency, showing the mechanical resonance used for height feedback. The upper and lower panels show the resonance peak at room temperature and 70 mK, respectively. 
}
\end{figure*}

Microwave imaging techniques have been developing rapidly over the past several decades, which involve transmitting MHz to GHz microwave signals to sample surface through sharp-ended probes, and collecting the microwave response. To date there are several implementations of the scanning microwave microscope, such as sharp tips in coaxial resonators, microwave resonators on quartz tuning forks, and probes with impedance matching networks \cite{Takeuchi1997YBaCuO, lee2005microwave, Graaf2013mK, graaf2015coherent, geaney2019near, lai2007atomic}. In this article we focus on the specific implementation of microwave impedance microscopy (MIM), which has shown capacities to probe the local conductivity and permittivity of quantum materials with nanoscale spatial resolution\cite{Rosner2002sum, lai2007atomic, Huber2010CalibratedNC, Lai2011NanoscaleMM, Lai2011, Kund2013unexpected,ma2015unexpected,ma2015mobile, liu2015thermal,seabron2016scanning, Chu2020,Barber2022}. 
This enables direct visualization of the microscopic nature of electronic states, including the real-space disorder landscape, multi-domain behavior, high-Tc superconductors, or the presence of topological modes that propagate along the sample boundaries\cite{Takeuchi1997YBaCuO, Lee2003YbaCuO, lee2005microwave, Tai2014Tc, graaf2015coherent, geaney2019near}.
By coupling microwaves with a wavelength of 1-\SI{100}{cm} to a sharp metallic probe and collecting the reflected signal, MIM characterizes  the complex admittance between the tip and the sample without the requirement for the sample to be conductive, which is less restrictive than other electronic imaging techniques\cite{eriksson1996cryocafm,doring2016pfm,lu2017stm,mcgilly2019pfm,rosenberger2020cafm}. 
As demonstrated in recent experiments, MIM can provide insight into the real-space nature of correlated states and topological states in two-dimensional heterostructures\cite{Lai2008, Lai2010, Ma2015, Wu2018,Allen2019, Chu2020, Barber2022}. However, many of these states are characterized by low energy scales and are therefore most robust at millikelvin temperatures, motivating the development of cryogenic MIM instrumentation.
Thus far, most state-of-the-art near-field microwave imaging has been performed in 1.5-$\SI{2}{K}$\cite{Worasom2011} or He-3 cryostats, which can reach of a minimum temperature of 300-\SI{450}{mK} \cite{Graaf2013mK, Allen2019}.

Here we report on the construction of a novel milliKelvin MIM, which will support spatially-resolved detection of quantum electronic states at ultra-low temperatures. This setup consists of scanning probe microscope with tuning-fork-based height feedback integrated into a dry dilution refrigerator. A sharp metallic probe driven by an AC signal at microwave frequency is coupled to the tuning fork and scanned over the sample. Using reflectometry, MIM detects the sample's response to high frequency electromagnetic fields emanating from the probe.

To demonstrate the measurement capabilities of this setup, we present MIM images of the conductivity contrast between graphite and SiO$_2$ down to temperatures of ~70 mK. 
Finally, we also demonstrate microwave imaging of edge states in \ce{Cd3As2} thin films in the quantum Hall regime at the base temperature.


\section{\label{sec:level1}Experimental Setup}

This setup consists of an custom-designed tuning fork based atomic force microscope (AFM) integrated into a Leiden Cryogenics CF-CS110 dilution refrigerator. The microscope housing is in thermal equilibrium with the mixing chamber plate on the cold-insertable probe, which is loaded into a dilution refrigerator, as shown schematically in Figure \ref{fig:schematics}(a). Figure \ref{fig:schematics}(b) shows the design of the microscope head, which houses an etched tungsten wire mounted onto to the end of one prong of a tuning fork (TF) mechanical resonator (blue box) \cite{Khan2012}. The oscillation amplitude of the TF is monitored for continuous height feedback, which enables tapping-mode topographic imaging\cite{Edwards1997}. Below the tip holder, the sample stage is mounted on a stack of CuBe piezoelectric scanners (Attocube AN-Sxyz100) and the positioners (ANPx(z)100), which control fine xyz scanning (up to $\SI{40}{\mu m}\times\SI{40}{\mu m}$ below $\SI{4}{K}$) and coarse positioning ($\SI{5}{mm}\times\SI{5}{mm}$ below $\SI{4}{K}$), respectively.

On the MIM circuitry side, GHz signals are generated by an analog signal generator, and one split branch of the signal is coupled to the tip via an impedance matching network (IMN) \cite{pozar2011microwave}, which is responsible for minimizing the reflected signal [inset in Figure \ref{fig:schematics}(c)] \cite{Cui2016}.  Unless otherwise mentioned, measurements in the article are done with \SI{0}{dBm} source power, and the power reaching the tip is \SI{-33}{dBm}.  A plot of the reflected microwave power $S_{11}$ of an example IMN is shown in Figure \ref{fig:schematics}(c), showing the first resonance at \SI{1.8}{GHz}.  The reflected signal from the tip passes through two directional couplers mounted on the probe-still plate ($\SI{1}{K}$) to cancel out the residual reflected power.

The signal from the sample is then amplified by a cryogenic amplifier (Cosmic Microwave Technology CITCRYO1-12) mounted on the $\SI{3}{K}$ stage, after which the signal propagates out of the probe and gets further amplified and demodulated at room temperature, as shown in Figure \ref{fig:schematics}(a).

During the tip approach procedure, active height feedback can be performed by monitoring either the TF oscillation amplitude or the MIM signal. Here we use a Nanonis SC5 controller to excite and track the TF oscillation and control the fine scanners during imaging \cite{Cui2016}.Figure \ref{fig:schematics}(d) displays a measurement of the oscillation amplitude of the tuning fork as a function of excitation frequency, showing the resonance peak near 32.768 kHz. 
The Q-factor of the resonance is around 500 - 2000 at room temperature (upper panel), while at base temperature it can easily reach 10,000-100,000 (lower panel).

The main technical challenge of microwave imaging in a dry dilution fridge is the emergence of new noise sources, which impact both spatial resolution and the signal-to-noise of the microwave reflectometry measurements. There are two main sources of increased noise: (1) mechanical pulse tube vibrations, which are associated with the cooling mechanism of the dilution fridge, place limits on the lateral spatial resolution and add noise to the measured MIM signal, and (2) the high Q factor of the tuning fork at mK temperatures leads to fluctuations in the tip-sample distance, which also couples with the pulse tube vibration.
Our fridge is equipped with a pulse tube cryocooler operating at $\sim \SI{1.4}{Hz}$ \cite{wang2002development,CHIJIOKE2010266} generating vibrations that amplitude modulate the tuning fork oscillation, and consequently also modulate the GHz MIM signal. 
To mitigate these vibrations, we physically decoupled the pulse tube motion from the microscope by unscrewing the rotary valve from the fridge and putting isolation foam in between \cite{li2005vibration}, while the high-pressure helium lines connected to the compressor are wrapped with acoustic pipe lagging. On the other hand, the high Q factor can be reduced by further breaking the symmetry of the two prongs, and two methods we have been using include adding extra glue during the tip-gluing process, and balancing the strain from the tip to the TF.

We found that performing AC-mode MIM imaging, described below, largely eliminates background oscillations in the images that arise from pulse tube vibrations.
In AC height-modulated imaging mode, a low frequency lock-in amplifier (SR830) is added to the output of the GHz frequency mixer to demodulate the reflected MIM signal at the tuning fork resonance frequency (32 kHz), after which low-pass filters can be used to attenuate noise \cite{Cui2016}.
We note that because the GHz MIM signal (from the tip) is amplitude-modulated by both the tuning fork oscillation at 32 kHz and the pulse tube vibration, there are multiple side-bands around the measurement frequency. 
Therefore, voltage preamplifiers with band-pass filters between 10-30 kHz (SR560, 6 dB/oct) are added to the output of the GHz mixer to both filter out the DC-MIM signal which would overload the lock-in amplifier, amplify the signal amplitude and reduce the out-of-band noise, after which the MIM signal is fed into the SR830 lock-in amplifier for demodulation.
During this step, the lock-in amplifier multiplies the MIM input signal with a TF reference signal (provided by the measured piezo current from the tuning fork, after amplification by a commercial low-noise charge amplifier) to extract the in-phase components.  
Both the filters inside SR830 and the additional low-pass filters  (SR560, 1 Hz low-pass) added to the output of the lock-in are chosen to eliminate noise at the pulse tube vibration frequency.


\begin{figure}
\includegraphics[width=0.48\textwidth]{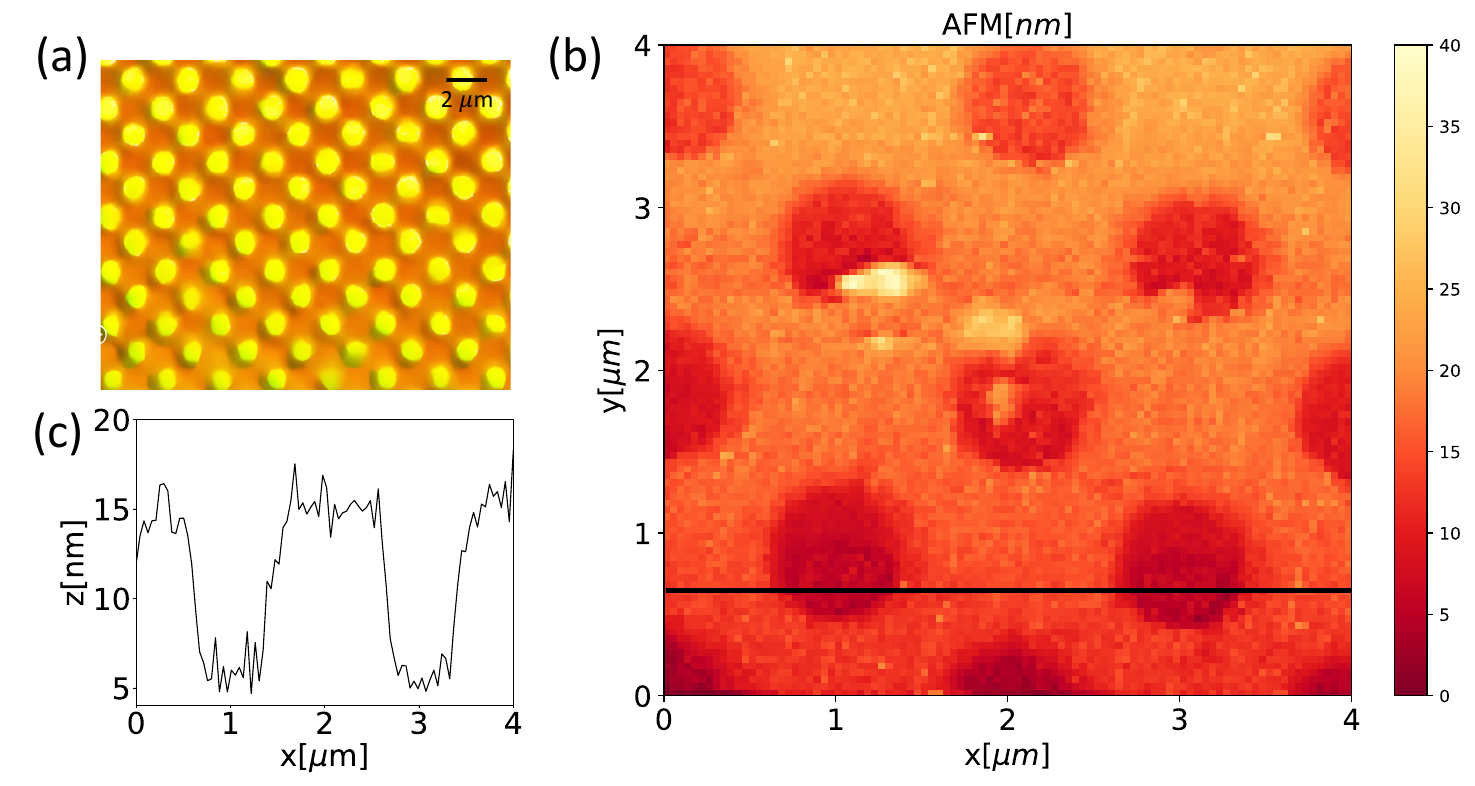}
\caption{\label{fig:afm} \textbf{Topographic imaging of a micropatterned dielectric film at mK temperatures using tuning-fork-based atomic force microscopy.} 
\textbf{(a)} Optical image of an etched array of holes $\mathrm{SiO_2}$. The diameter and the spacing of the holes are 1 $um$. The hole depth is \SI{20}{nm}. 
\textbf{(b)} AFM spatial scan at 70mK. The scan covers 4$\times$ \SI{4}{um} and the scan speed is  400 $nm/s$.
\textbf{(c)} Cross-sectional line cut corresponding to the black line in (b).}
\end{figure}



\begin{figure*}
\includegraphics[width= 0.85\textwidth]{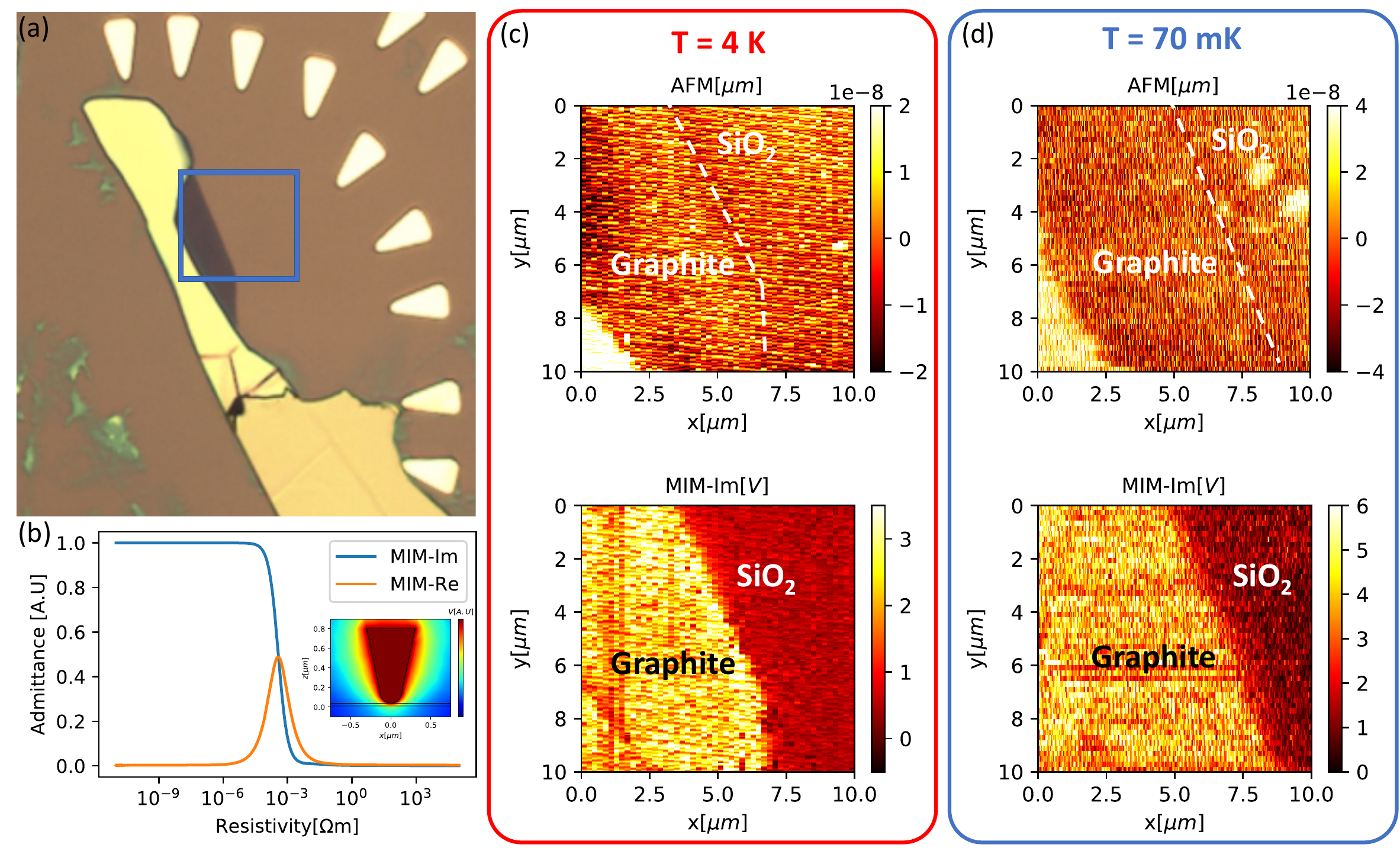}
\caption{\label{fig:graphite2} \textbf{Microwave impedance microscopy of graphite at milliKelvin temperatures.}
\textbf{(a)} Optical image of a graphite flake exfoliated onto a SiO$_2$/Si substrate. The dark purple region has a thickness of 3-\SI{5}{nm}, and the light yellow region has a thickness of $\sim\SI{20}{nm}$. The blue box marks the imaging window for (c) and (d). 
\textbf{(b)} Theoretical MIM response curves simulated at \SI{1.8}{GHz}, illustrating the evolution of the MIM contrast with the sample conductivity. \textit{Inset:} vertical cut of the potential distribution for the tip-sample interaction, calculated using finite-element analysis. 
\textbf{(c)} AFM and MIM imaging of the graphite flake at $\SI{4}{K}$, with the scan window covering the \SI{20}{nm} region (lower left), 3-\SI{5}{nm} region (middle), and the SiO$_2$ region (upper right). The scan speed is 0.5 $\mu m/s$ . 
\textbf{(d)} AFM and MIM images of the same location at $\SI{70} {mK}$. The scan speed is 0.2 $\mu m/s$.} 
\end{figure*}



\section{\label{sec:level1}Results and Discussion}

We characterized the low temperature performance of the AFM on a sample consisting of an array of etched  $\mathrm{SiO_2}$ holes patterned on a Si wafer, as depicted in the optical image in Figure \ref{fig:afm}(a). 
Cryogenic AFM measurements are used to visualize the topographic profile of a 5 $\mu$m $\times$ 5 $\mu$m scan region at $\SI{70}{mK}$, as depicted in Figure \ref{fig:afm}(b). 
Figure \ref{fig:afm}(c) shows a cross-sectional cut of this AFM image, whose position is marked by the black line, revealing a noise level of roughly \SI{3}{nm}. 
To more carefully assess the magnitude of the z-noise during AFM scanning, we performed 96 $\times$ 96-pixel noise scans over a 1 nm $\times$ 1 nm area, such that the spatial profile is irrelevant. Root mean square (RMS) roughness was calculated using Gwyddion after line fitting, which gives z-noise levels in the range of 1.8 - 2.2 nm. 
Furthermore, upon careful inspection of Figure \ref{fig:afm}(b), we noticed that a tilted stripe pattern appears as a background modulation in the AFM image. By taking a Fourier transform of this data, we found that the stripe pattern has a frequency of \SI{1.4}{Hz}, which coincides with the frequency of the pulse tube.

Next, to demonstrate our combined AFM and MIM imaging capabilities at low temperatures, we measured the spatial contrast of the MIM response across the boundary between graphite and SiO$_2$ at 70 mK. 
Figure \ref{fig:graphite2}(a) shows an optical image of the graphite sample, which has terraces of varying thicknesses: the purple region is $\sim\SI{3}{nm}$ and the bright yellow region is 15-\SI{20}{nm}. 

In Figure \ref{fig:graphite2}, panels (c) and (d) display AFM and MIM images of the graphite/SiO$_2$ interface measured at 4K and \SI{70}{mK}, respectively.
In both sets of AFM images, the 3/\SI{20}{nm} step height 
in graphite is clearly visible, while the graphite/SiO$_2$ boundary only shows a faint contour, as the z-movement of the scanner to compensate for fluctuations in the tip-sample distance dominates over the \SI{3}{nm} boundary. 
Meanwhile, we observe a sharp contrast in the MIM signal across the graphite/SiO$_2$ boundary due to the different conductivities of the two materials, as predicted by the response curves in Figure \ref{fig:graphite2}(b). 
To explain the experimental observations, one can model the tip-sample interaction for this system using finite element analysis\cite{Wu2018}. In the numerical simulation, the tip is defined to be \SI{1}{V} and the bottom of the \ce{SiO2} is set at \SI{0}{V}, the ground. The side of the sample is electrically floating. For a sample with a scalar conductivity, the simulation can be used to calculate the MIM response curves as a function of sample resistivity, which should be expected to have a response curve similar to the lump-circuit model defined for MIM imaging\cite{Barber2022,EMPtheory}. For a better discussion of lump-circuit model, readers can refer to the supplementary information. At a measurement frequency of 1.8 GHz, the imaginary part of the MIM response should monotonically decrease with the sample resistivity, saturating when the resistivity is higher than $10^{-2}$ $ \Omega\cdot m$  (insulating limit) or lower than $10^{-5}$ $\Omega \cdot m$ (conductive limit), as shown in Figure \ref{fig:graphite2}(b). A cross-sectional profile of the penetration of the tip potential into the sample is provided in the inset. 
Based on our observation in the imaging, we estimate the MIM spatial resolution to be below \SI{200}{nm}, constrained by the apex geometry of the etched tungsten tip and mechanical noise from pulse tube vibrations.


\begin{figure*}
\includegraphics[width= 0.85\textwidth]{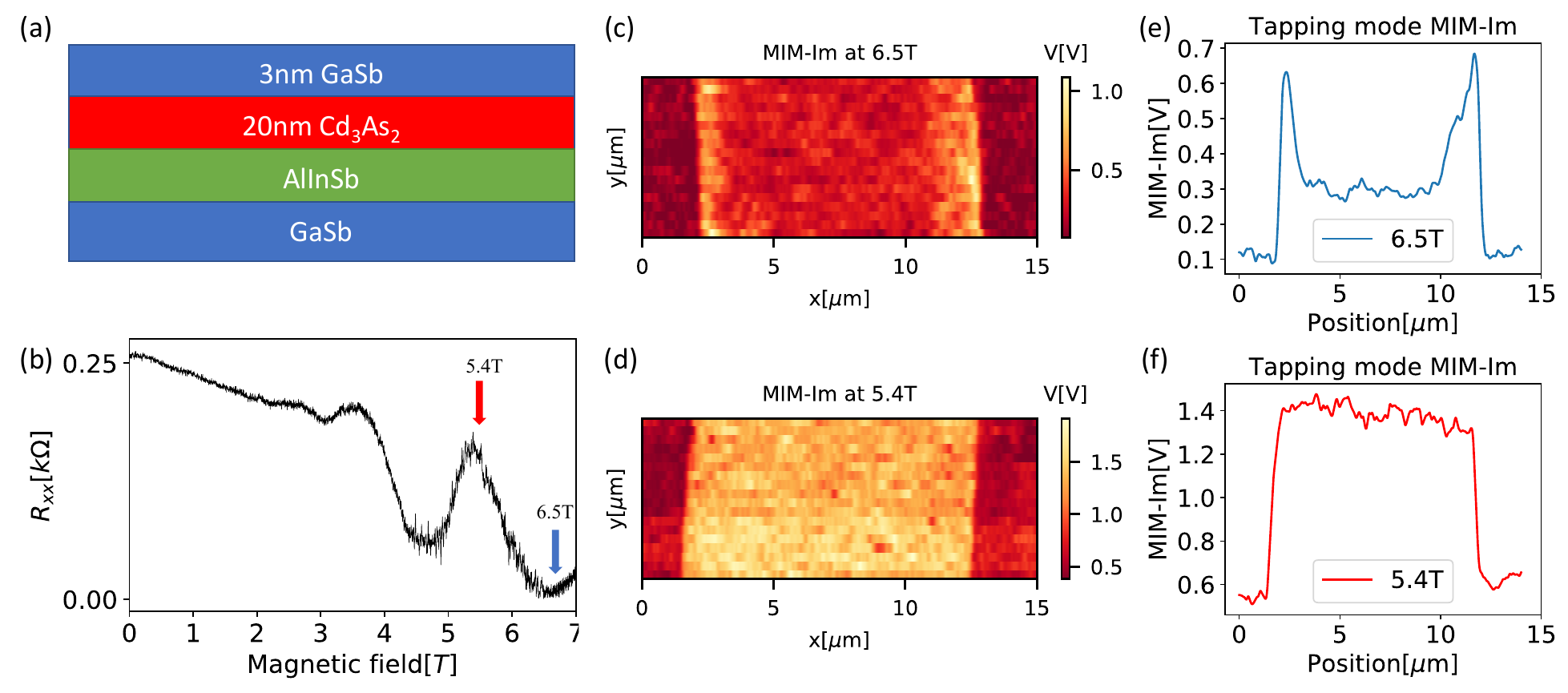}
\caption{\label{fig:CdAs} \textbf{Microwave imaging of edge modes in a cadmium arsenide film in the quantum Hall regime.} 
\textbf{(a)} Cross-sectional schematic of an epitaxially-grown  \ce{Cd3As2} heterostructure. 
\textbf{(b)} Transport measurement of the longitudinal resistance $R_{xx}$ as a function of magnetic field at \SI{90}{mK}. The minima correspond to the emergence of quantum Hall plateaus.
\textbf{(c)} MIM image at \SI{6.5}{T}, revealing a sharp enhancement of the reflected signal at the boundaries of a quantum Hall insulator state.
\textbf{(d)} MIM image at \SI{5.4}{T}, showing spatially uniform conductivity at the transition between quantum Hall plateaus
\textbf{(e-f)} Cross-sectional line cuts of the MIM response across the sample extracted from (c) and (d) respectively.
}
\end{figure*}


We also apply this methodology to visualize edge states at the boundaries of thin film cadmium arsenide (\ce{Cd3As2}), a novel three-dimensional Dirac semi-metal, in the quantum Hall regime\cite{sankar2015Cd3As2,schumann2016MBE}. A cross-sectional schematic of the epitaxially-grown hetero-structure is shown in Figure\ref{fig:CdAs} (a), where the film thickness is \SI{20}{nm}\cite{goyal2018thickness,lygo2023two}. The \ce{Cd3As2} device is lithographically patterned and etched into strips of width 10-15 $\mu$m, which are electrically grounded. 
Transport measurements were performed to characterize the magnetic field behavior of the sample, which reveal dips in the longitudinal resistance at around \SI{4.7}{T} and \SI{6.5}{T}, as shown in Figure \ref{fig:CdAs}(b).  These minima should correspond to the emergence of quantum Hall plateaus\cite{guo2022nu}.

To shed light on the real-space conductivity profile of \ce{Cd3As2} in the quantum Hall regime and monitor its evolution across the topological phase transition between plateaus, MIM measurements were performed at a series of magnetic fields at a base temperature of 90 mK.   
Microwave imaging reveals a sharp enhancement of the reflected MIM signal at the boundaries of the sample in the quantum Hall insulator state, which rapidly decays into the bulk of the sample, as shown in Figure \ref{fig:CdAs}(c).
Meanwhile, we observed a spatially-uniform conductivity at the transition between quantum Hall plateaus when the longitudinal resistance deviates from zero at B = 5.4 T (Figure \ref{fig:CdAs}(d)). The variation of the MIM signal between different lines comes both from the noise in the MIM signal and spatial inhomogeneities in the sample.

To more clearly compare the spatial dependence of the MIM signal in these two regimes, in Figure \ref{fig:CdAs}(e-f) we plot the cross-sectional profiles of the MIM response across the sample extracted from panels (c) and (d), respectively.
These low temperature microwave images reveal sharply enhanced edge conduction that encloses an insulating  interior in the quantum Hall regime, which is consistent with the results of transport measurements performed on this system in prior experimental studies.

\begin{figure}[!h]
\includegraphics[width=0.48\textwidth]{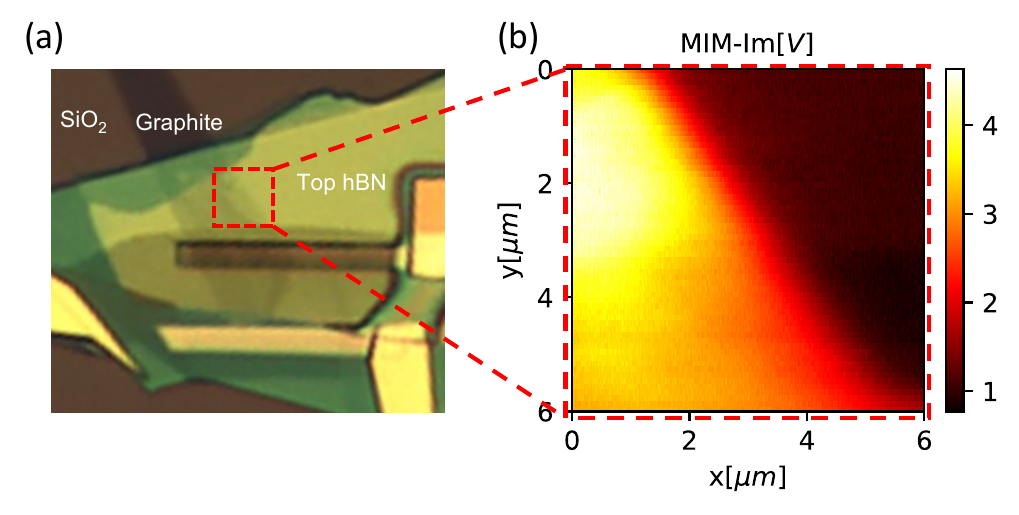}
\caption{\label{fig:Floating} \textbf{Demonstration of low-noise height-modulated MIM in a dry dilution fridge. } 
\textbf{(a)} Optical image of a van der Waals heterostucture with a graphite flake (around \SI{5}{nm} thick) covered by hBN (around \SI{50}{nm} thick). From bottom up the whole stack is graphite (5 nm, connected from the leftmost gold contact), bottom hBN(30 nm, dark green), graphite contact(5-10 nm, middle) and top hBN (50 nm, bright green). The imaging window has 30 nm hBN, 5-10 nm graphite and 50 nm hBN from bottom up.
\textbf{(b)} A floating mode MIM image of the region enclosed by the red square in panel (a), acquired at $\SI{70}{mK}$.
The measurement frequency is $\SI{3}{GHz}$ and the tip is retracted $\SI{100}{nm}$ from the highest feature inside the scan window.
}
\end{figure}

We note that one way to improve signal quality is to use ``floating'' AC-mode MIM, where imaging is performed with the tip retracted a fixed distance (60-$\SI{100}{nm}$) above the sample surface.
At this distance, the AFM channel will not be modulated due to the topography feedback, but the MIM tip can still interact with the sample via the electromagnetic fields in the vicinity of the tip (when operated in the near-field regime). Because periodic oscillations in the tip-sample distance at the tuning fork resonance are decoupled from the surface roughness of the sample, noise in the MIM response can be dramatically reduced in floating mode. Figure \ref{fig:Floating} shows the results of a floating mode MIM scan performed at $\SI{3}{GHz}$ and T= $\SI{70}{mK}$, with the tip lifted $\SI{100}{nm}$ above an hBN-covered graphite layer. The tip apex is around $\SI{0.8}{um}$, which is reflected in the spatial profile of the MIM signal change across the boundary between the graphite flake and hBN. In this case, the signal-to-noise ratio is even better than that observed in tapping mode MIM images (Figure \ref{fig:graphite2}(c-d), which is especially useful for fixed-location MIM measurements. However, this advantage comes at the expense of signal size, as the tip is further away from the sample than for tapping mode. The varying MIM signal in the graphite region is due to the fact that during a floating scan the tip is hovering at a fixed absolute z-position, while the sample has a surface tilt, which means the tip-sample distance is varying in the whole frame. We estimate the distance to be around \SI{70}{nm} in the upper-left corner and \SI{100}{nm} in the lower-right corner. 

The choice of tip-sample distance for floating-mode measurements is a compromise between maximizing signal sensitivity and minimizing the risk of a tip crash due to vertical fluctuations in the tip-sample distance, which arise from pulse tube vibrations and are aggravated by the large Q factor of the tuning fork at mK temperatures. For larger scan windows or rougher sample surfaces, the tip may need to be retracted further. We expect the sensitivity of floating mode to be around $0.01-\SI{0.1}{aF/\sqrt{Hz}}$ at \SI{0.1}{uW} input power, and in our case the noise is mostly due to vertical modulations of the tip-sample distance\cite{Lai2008}.


\section{Conclusion and outlook}
In summary, we report on the development of a microwave impedance microscope that operates at temperatures down to \SI{70}{mK}. This is achieved by incorporating a TF-based AFM with near-field GHz imaging capabilities into a dry dilution refrigerator. 

Pushing the limits of MIM into new low temperature regimes should enable
local sensing of quantum phenomena that only exist at low energy scales, including certain topological states of matter, domain wall physics at phase transitions, quantum states arising from geometric confinement in mesoscopic devices, and correlated states in two-dimensional materials and van der Waals heterostructures. 
Because this instrumentation is equipped with combined transport and imaging capabilities, it can also illuminate the correspondence between macroscopic transport behavior and the underlying microscopic nature of electronic states, including the real-space disorder landscape or presence of edge modes.

During the preparation of this manuscript, we became aware of an article on a related topic\cite{KejiLai2023implementing}. We are using the same tuning-fork based MIM circuitry, and in this manuscript we focus our efforts on mitigating the relevant AFM and MIM noise caused by the fluctuation in tip-sample distance. The z-resolution of AFM in the reference is in the order of 0.1-0.3 nm, while in our case it’s 3-20nm. The MIM sensitivity in the reference is mainly limited by the MIM setup, while our MIM sensitivity is constrained by the vertical tip-sample variation. 

\section*{supplementary material}

The supplementary material includes more information on the analysis of the pulse tube vibration frequency, a description of the lumped element model of MIM, and the determination of the MIM spatial resolution.

\begin{acknowledgments}
We thank Alex Lygo and Susanne Stemmer for providing cadmium arsenide devices for these experiments, Yongtao Cui for inspiring discussions, and Evan Cobb for helping develop some of the MIM circuitry.  
We gratefully acknowledge funding support from the UC Office of the President, specifically the UC Laboratory Fees Research Program (award LFR-20-653926), the AFOSR Young Investigator Program (award FA9550-20-1-0035) and the AFOSR/ARO MURI Program (award FA9550-22-1-0270). This work was performed, in part, at the San Diego Nanotechnology Infrastructure (SDNI) of UCSD, a member of the National Nanotechnology Coordinated Infrastructure, which is supported by the National Science Foundation (Grant ECCS-2025752).

\end{acknowledgments}

\section*{Data Availability}

The data that support the findings of this study are available from the corresponding author upon reasonable request.

\nocite{*}
\bibliography{aipsamp}

\end{document}